\documentclass{cimento}

\usepackage{graphicx}
\usepackage{enumitem}

\title{A high-level characterisation and generalisation of
communication-avoiding programming techniques}
\shorttitle{High-level characterisation and generalisation of communication-avoiding} 
\author{T.~Weinzierl
  \from{ins:DurhamCS}
  \from{ins:DurhamIDAS}
}
\instlist{
  \inst{ins:DurhamCS}
  Department of Computer Science,
  Durham University, 
  Durham,
  Great Britain
  \inst{ins:DurhamIDAS}
  Institute for Data Science,
  Durham University, 
  Durham,
  Great Britain
}

\begin{document}

\maketitle

\begin{abstract}
  Today's hardware's explosion of concurrency plus the
explosion of data we build upon in both machine learning and scientific simulations have multifaceted
impact on how we write our codes.
They have changed our notion of performance and, hence,
of what a good code is:
Good code has, first of all, to be able to exploit the unprecedented levels of
parallelism.
To do so, it has to
manage to move the compute data into the compute facilities on time.
As communication and memory bandwidth cannot keep pace with the growth in
compute capabilities and as latency increases---at least relative to what the
hardware could do---communication-avoiding techniques gain
importance.
We characterise and classify the field of communication-avoiding algorithms.
A review of some examples of communication-avoiding programming by means of our
new terminology shows that we are well-advised to broaden our notion of
`communication-avoiding'' and to look beyond numerical linear algebra. 
An abstraction, generalisation and weakening of the term
enriches our toolset of how to tackle the data movement challenges.
Through this, we eventually gain access to a richer set of tools that we can use
to deliver proper code for current and upcoming hardware generations.

\end{abstract}

\section{Introduction}

Advance in computational sciences, engineering, medicine and other
computational disciplines often relies on simulation results to drop in faster,
even though researchers continuously increase the problem size or detail.
There are other important driving factors behind progress such as improved
models or numerical techniques.
Yet, speed and size are omnipresent desirable characteristics and
requirements.
To suit them, the community relies on the improvement of two tools:
their computer hardware and their software.
Improvement here means speed.

Computers are becoming faster.
The Top-500 list \cite{Webpage:Top500} documents their impressive evolution.
In many manuscripts, the speed growth is attributed towards Moore's law; which
is not in line with the original statement from \cite{Moore:65:Law}.
What is however true is that the growth---though maybe with a reduced constant
before the exponential---of Moore's law does and continues to hold for
computers' potential capabilities.

Parallelism is key to make computers faster
\cite{Asanovic:2006:LandscapeOfParallelComputing}.
Starting from around 2004, computers stagnated in frequency or even
reduced it \cite{Sutter:05:FreeLunchIsOver}. 
This is due to physical (quantum) effects introducing nondeterministic
effects, i.e.~errors in a deterministic view of the world, energy
envelopes, manufacturing constraints, and so forth.
All effectively forbid
vendors to increase the frequency or to shrink the dies aggressively.
Shrinking allows us to work at lower voltages which in turn
creates freedom to increase the speed.
Once these upscaling techniques are not an option anymore, we can still
increase the number of compute units. 
This yields more powerful computers despite the fact that individual compute
components do not speed up.

The seminal, popular science report ``The Free Lunch Is Over: A Fundamental Turn
Toward Concurrency in Software'' \cite{Sutter:05:FreeLunchIsOver} documents this
insight as well as its consequences:
software has to be made ready for parallelism, too.
A second seminal paper is the SCaLeS report championed by David E.~Keyes
\cite[pp.~32]{Scales:03:Scales} which contains content similar to an algorithmic version of Moore's Law:
The report highlights that the dramatic growth of performance and capabilities
on the hardware side over the last decades is matched by improvements on the
algorithm side.
The manuscript's introductory example from linear algebra showcases this by the
transition from direct Gauss Elimination into full multigrid \cite{Trottenberg:01:Multigrid} which
yields a gain in compute efficiency equivalent to 35 years of hardware evolution.
More examples along these lines exist.

If we cast statements on algorithmic improvements into statements on performance
without machine context, we run risk to make an error very similar to the
misinterpretation of Moore's law as performance statement.
Improvements on the hardware side and on the algorithm side are not 
orthogonal:
In the present paper, we sketch nuances of the growth of hardware parallelism
and highlight that the sole focus on an increase of concurrency\footnote{
 In the tradition of \cite{Lamport:78:Concurrency}, we use concurrency as a
 theoretical concept that highlights that two control flows in hardware or in
 code run independently, can be merged into each other in any order, or may be
 executed in parallel, i.e.~it is the term parallelism that denotes that things
 happen at the same time.
} 
on the software side falls short, as data movements quickly become a
limiting factor (Section~\ref{section:co-design}).
As a consequence, programmers of high-end software have to familiarise
themselves with communication-avoiding techniques.
In Section~\ref{section:terminology}, we attempt to classify
communication-avoiding techniques and claim that the phrase has to be weakened
and generalised to include many more techniques than the traditional
communication-avoiding tool set from linear algebra.
Consult \cite{Demmel:2013:CA} for a comprehensive overview over the latter.
Only a generalisation and, hence, an expansion of our toolset 
accommodates the pressing needs imposed on us by the hardware
evolution.
The two sections \ref{section:strong} and \ref{section:weak} give examples for
communication-avoiding techniques in line with this generalised notion.
The examples allow us to conclude (Section~\ref{section:conclusion})
which next steps might be required to tackle the
challenges around communication in data-intense computing.

\section{One-way co-design}
\label{section:co-design}

%
%
Reports and roadmaps predict that the exascale wall will be climbed
through an increase of parallelism on the hardware side
\cite{Asanovic:2006:LandscapeOfParallelComputing,Dongarra:14:ApplMathExascaleComputing}.
There are different levels of parallelism.

First, vendors increase the parallelism through the widening of vector
registers---AVX512 is an example---or lockstepped hardware threads as we find
them in GPGPUs.
Parallelism on this level materialises in code which performs basic arithmetic
operations on small, fully populated vectors or vector segments.
NVIDIA calls its cores streaming multiprocessors (SMs) which work with many
sets (warps) of hardware threads per core which are swapped quickly.
This original SIMT's (single instruction multiple threads)
\cite{Lindholm:2008:CUDA} concept is conceptionally
close to vectorisation.
Its swapping idea materialises hardware concurrency hiding
latency and bandwidth constraints.

Second, vendors increase the core count per processor.
Intel might have abandoned their own manycore product line, but we see
mainstream supercomputers nevertheless host already 20 or more cores
per processor.
Different to vector registers, multicores run independently of each
other, i.e.~each executes its own instructions. 
GPGPUs in general operate with lower physical core/SM counts.
As NVIDIA however supports higher thread divergence, i.e.~allows multiple
threads ran simultaneously to run different code branches and to synchronise
with a subset of other threads, the SIMT paradigm moves closer to multicore
parallelism.
Cores and SMs are connected to each other through a shared memory.
One core thus can read data written by another core.

Finally, supercomputers connect multiple of these multicore devices (nodes) with
each other.
Nodes do not share data spaces but communicate with each other
through messages.
Alternatively, new hardware in the market, such as NVLink, 
joins distributed memory logically into large shared spaces and thus pushes
multi-node or multi-device architectures logically into the multicore domain.

Recent hardware generations exhibit a fast growth of concurrency on the vector
and multicore level, while the number of nodes seems to stagnate.
One reason for this are data movements which are subject of the subsequent
sections: 
If it is already expensive and problematic to transfer data between
memory and vector registers as well as in-between cores, it is even more
challenging to transfer data over long distances via messages.

%
%
In a co-design fashion, the tremdendous increase in concurrency on the hardware
side forces developers to rewrite their codes such that they are capabable of
exploiting this \cite{Sutter:05:FreeLunchIsOver}. 
The different hardware levels here require different programming efforts.
An increase in vectorisation parallelism forces programmers to phrase
algorithms in terms of small vectors.
These vectors are treated as dense, i.e.~vector units on a chip manipulate vectors agnostic of their content.
An increase in core counts requires programmers to decompose algorithms into
fine-granular activities (tasks). 
This allows a runtime system to deploy tasks among the cores and, hence,
to keep them busy.
An increase in node counts requires programmers to decompose data spaces into
larger regions and to design codes such that the effort to keep these
distributed data spaces consistent remains limited.
If we formalise the induced communication as graph, the graph has to be as
sparse as possible.
In linear algebra, this corresponds to the need for sparse matrices.

%
%
Co-design the other way round is harder to find. 
Algorithms rarely guide the hardware design \cite{Russell:2016:CoDesign}, and special-purpose
chips materialising particular algorithms are not found that often in mainstream
HPC.
Recently however, we witness an influence the other way round:
In-hardware security algorithms (encryption) become part of chips,
FPGAs---software-programm\-able hardware---gain momentum, and GPUs start to
become equipped with special-purpose silicon; 
tensor cores and raycasting/-tracing entities are the prime examples.
The economic success of GPGPUs as a whole has, to some degree, to be attributed
towards machine learning algorithms.
One may interpret this as algorithm-to-hardware co-design.
We however prefer a different interpretation:

\subsection{Changing optimality metrics: over-vectorisation and task
decomposition}
Many sophisticated, efficient algorithms reduce
the concurrency:
In traditional computer science, 
an algorithm is efficient if it requires fewer computations than a na\"ive,
competing approach. 
Many computations---preferable of the same type---however tend to yield a high
concurrency, while very few calculations, in the ``worst case'' with tight
data dependecies, cannot make use of a parallel machine.
Sometimes, simpler is faster.

Some sophisticated codes thus combine the best algorithm on one scale with a
simpler variant with higher concurrency on another scale.
Examples for such hybrids that support our efficiency-vs-concurrency claim are
block-structured sparse matrix algorithms that use full matrix algorithms for subblocks
\cite{Ltaeif:19:HierarchicalAlgorithms}, 
linear algebra subroutines which switch to dense or sparse implementations
depending on the number of non-zeros and their pattern \cite{Breuer:2014:SustainedPetflow},
particle dynamics codes that use a grid
to administer the cut-off radius which is so coarse that it always hosts a
reasonable number of molecules
\cite{Eckhardt:15:SPHCompression,Schaller:16:Swift,Weinzierl:15:PIC}, or
multigrid codes that rely on algebraic multigrid for coarse resolutions and
employ geometric multigrid with rediscretisation on finer levels
\cite{Rudi:15:ExtremeScaleImplicitSolver,Weinzierl:17:BoxMG}.
Neural networks often are an extreme case:
Conceptually, they are significantly simpler than known approaches to minimise
cost functions describing how well a network represents a training or test data
set, respectively.
Their success in fitting strongly
overparameterised ansatz functions stems from massive data plus unprecedented concurrency; typically
subject to gradient descent, i.e.~a relatively basic algorithmic approach.

Besides the aforementioned revisit of ``less effective'' algorithms
which allow us to exploit (vector) concurrency and thus effectively
reduce the time-to-solution, the decomposition of our algorithms into finer and finer
tasks is the second important code design trend in large-scale computing.
The latter technique uncovers fine-granular concurrency and
is often followed by a rearrangement of these tasks such that the machine of choice is exploited
better. 
The rise of task systems---Threading Building Blocks, Cilk, Charm++,
StarPU, and others---documents the popularity of the latter approach.
Developers have to carefully balance task decomposition against vectorisation
efficiency: 
Too small tasks fail to keep the vectorising codes busy.
Too big tasks lack the required level of concurrency.
Both dimensions benefit from homogeneous, non-hierarchical, uniform
workload which can be split relatively straightforwardly
\cite{Asanovic:2006:LandscapeOfParallelComputing}.
Eventually, both benefit from hybrids.

The paradigm shift to construct hybrids exploiting less
efficient algorithms than theoretically known results from an update of our cost
model, i.e.~our understanding what induces high or low runtimes respectively.
New cost models are rarely introduced explicitly.
Most of the time we alter them implicitly in our minds and support the outcomes
through experimental studies.
Yet, the punchline remains the same: 
The delicate balance between high concurrency vs.~parallelisation overhead
in some cases outweights classic operation count-based cost models.
\emph{The paradigm shift on the hardware side has recalibrated and changed the
correlation of floating point operation counts (FLOP numbers) to performance.}

\subsection{Vertical and horizonal communication constraints}
%
%
A second challenge rises in the shadow of hardware concurrency growth:
The architectures become ill-balanced w.r.t.~data delivery.
While the cores with their vector registers can yield an impressive number of
computations per second and while there are many cores, we struggle to feed them
with data
\cite{Dongarra:14:ApplMathExascaleComputing,Higham:2017:News,Higham:2019:SqueezingMatrix,Laguna:2019:GPUMixer,Lam:2016:ShadowValueAnalysis,Langou:2006:IterativeRefinement}.

\begin{figure}
 \begin{center}
  \includegraphics[width=0.6\textwidth]{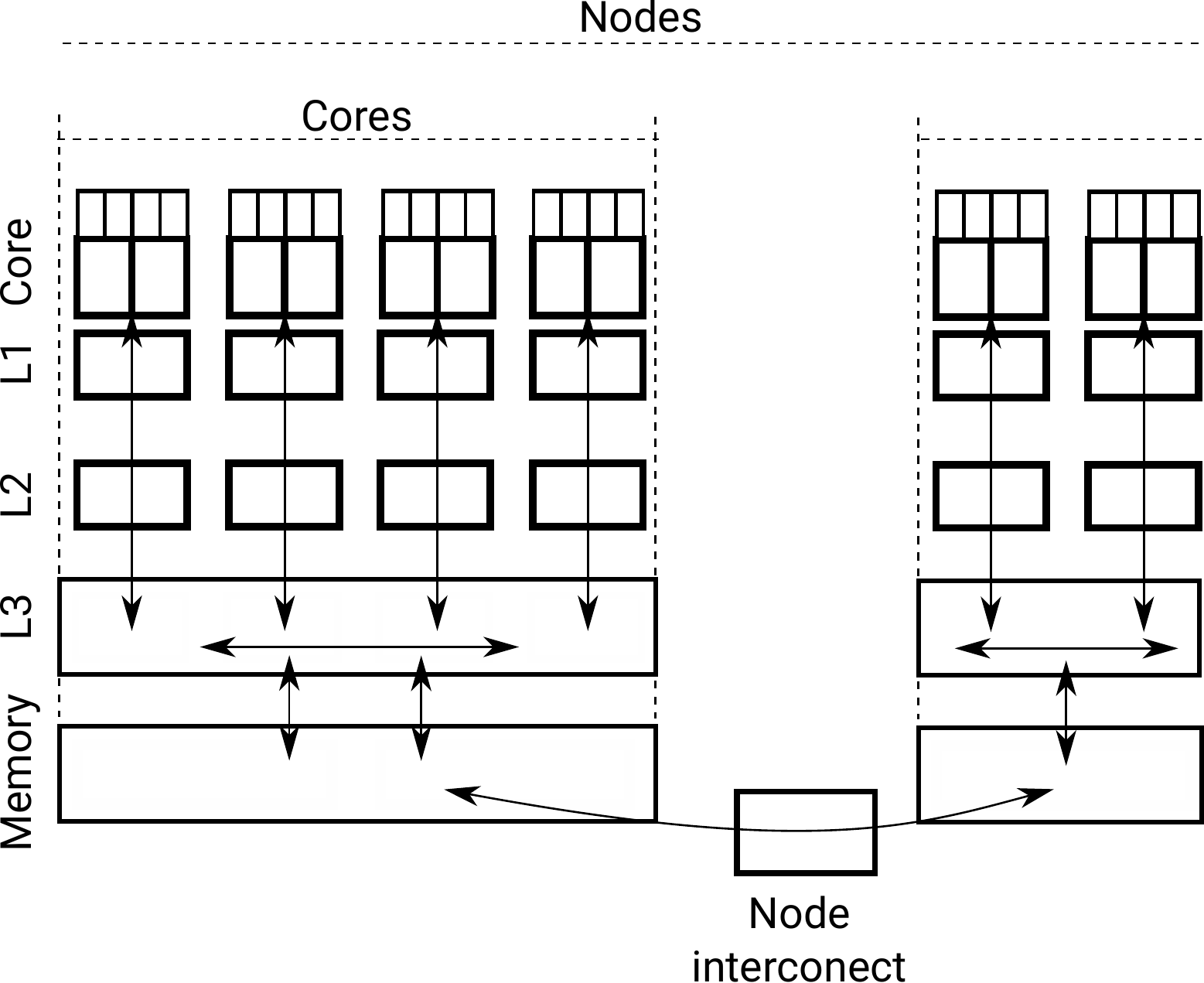}
 \end{center}
 \vspace{-0.4cm}
 \caption{
  Schematic illustration of vertical and horizontal data flow in a conventional
  supercomputer consisting of multiple nodes where each node hosts multicore cores which in turn employ
  vector registers. The cores access the main memory through a cascade of cache
  levels, while the nodes communicate with each other through an interconnect.
  \label{figure:vertical-and-horizontal-data-flow}
 }
\end{figure}

There is a \emph{vertical data movement challenge}
(Fig.~\ref{figure:vertical-and-horizontal-data-flow}):
Cores can process data which is by a magnitude larger than what a standard main
memory can deliver.
One hardware workaround here is to introduce caches, i.e.~intermediate memories
in-between the main memory and the compute units.
They are small yet can hold hot data, i.e.~data that is frequently or soon used.
An alternative to caching is the use of massive oversubscription:
Whenever a core runs into a situation where the computations of one
thread require data which is not yet brought into the chip's registers, it
``suspends'' the execution and instead progresses another ``warp'' of (vector)
operations from a different thread.
In the meantime, the required data can be brought in.
The latter technique is used in GPGPUs.
As sketched above, such chips logically can be interpreted as a set of cores
where only one core is active (and thus requires compute resources) at a
time, while the others prepare for their next move.

Intel introduces a new cache level with its Optane technology
\cite{Boyandin:18:OptaneInMemoryDB,Charrier:19:EnergyAndDeepMemory,Kudryavtsev:17:NUMAOptane}.
Optane effectively renders the main memory into a further cache---either
implicitly managed by the company or explicitly managed by user codes---and
introduces a huge yet relatively slow main memory.
Intel furthermore abandons the idea of inclusive caching with the 
Intel\textregistered\ Xeon\textregistered\ Scalable processor (Skylake).
Without inclusive caching, the probability increases that cache misses on a
small, close cache induce large runtime penalties: the missed data
is not backed up by the subsequent cache level.
If caches become (relatively) slower, it hurts even more.
On the GPGPU side, NVIDIA weakens the strict lockstepping in warps.
This allows users to run codes with a larger thread divergence, yet harms the
GPGPU's efficiency if used heavily.
The vertical challenge gains weight.

There is an additional \emph{horizontal dimension to the data movement
challenge} (Fig.~\ref{figure:vertical-and-horizontal-data-flow}):
Data movement penalties gain weight as vendors integrate more and
more cores into one chip.
Cache levels have to be split or shared between more and more compute
units and to keep their content consistent requires increasing effort
manifesting in runtime penalties (NUMA).
On the inter-node level, interconnects remain to be by magnitudes slower than
memory.
This also holds for processor-accelerator connections.

Both vertical and horizontal data movement constraints imply that efficiency
gains resulting from over-vectorisation and task decomposition run risk to
become obsolete.
The fastest compute instruction sequence, i.e.~source code, is of no value if
the processing units can not be fed with data.
The hardware evolution makes it mandatory that codes avoid communication along
all hardware dimensions.

\section{Terminology}
\label{section:terminology}

Communication-avoiding (CA) algorithms are not new.
One of the most important papers stems from 2013, e.g.,
\cite{Demmel:2013:CA} and reviews a vast number of CA techniques from
linear algebra.
We identify here six dimensions along which we can classify CA techniques:

\begin{enumerate}[leftmargin=*]
  \item 
In the tradition of refactoring \cite{Fowler:99:Refactoring}, 
communication-avoiding modifications always have to be semantics-preserving,
i.e.~the outcome of an algorithm subject to a communication-avoiding technique has to
remain the same. 
In scientific computing, ``the same'' usually means ``reasonably close to
machine precision''.
This implies that a communication-avoiding technique may not introduce
any numerical instability for the input data of interest, but we may classify
techniques into {\bf stability-preserving} and {\bf stable for data of
interest}.
  \item 
We consider both data movements in-between nodes and within a
computer, i.e.~between chip and
caches, caches and main memory, memory and accelerator, as communication.
This is, our notion of communication comprises {\bf intra-node and inter-node
communication}.
  The phrase communication-avoiding thus includes cache optimisation techniques
  \cite{Kowarschik:03:CacheTechniquesOverview}.
  As novel hardware blurs the distinction between intra-node and inter-node
  (through remote memory access or distributed shared memory for example), we
  may characterise techniques rather by means whether they address problems of
  {\bf horizontal or vertical information flow}:
Horizontal circumscribes challenges that arise from the growth of
concurrency such as increasing node and core counts but also wider vector
registers.
Vertical techniques tackle challenges that arise from the fact that information
typically is spread out, i.e.~resides on different memory levels,
memory segments, devices.
  \item 
Though papers traditionally motivate a communication-avoiding technique through
one application and demonstrate the impact of their ideas by means of one
application, we can distinguish {\bf application-specific from generic}
approaches. The latter techniques apply directly to many different areas.
  \item 
  There are two extremes of CA code changes:
  Techniques can {\bf preserve the exact algorithmic steps per datum}.
  If we work in
  this regime, we are restricted to a re-orchestration of operations over all data, but 
  the path how this outcome had
  been computed is preserved per datum. We are notably not allowed to change the
  algorithm that led to a particular result. 
On the other side of the spectrum are techniques and algorithms which
{\bf change the algorithm's behaviour} or even the algorithm itself, i.e.~they
change how results are obtained.
There are many approaches in-between; in particular approaches which weakly
preserve the behaviour, i.e.~yield different pathways towards the results only
in some cases.
The in-the-middle regime comprises for example code modififcations
which materialise in altered iteration counts, required operations, and so
forth but stick to the same algorithmic blueprint.
  \item Some CA techniques are {\bf strictly monotonous} in a sense that they
  always reduce the communication. Others pay off asymptotically for large,
  characteristic data sets yet not for each individual choice.
  \item 
Finally, many (application) scientists are not motivated by
communication-avoiding per se, but start to investigate CA techniques as
they suffer from communication penalties.
  From their point of view, it makes sense to 
  distinguish {\bf strong communication-avoiding} techniques from
  {\bf weak communication avoiding} techniques.
  Strong means that communication is eliminated compared to a na\"ive baseline
  implementation.
  In the weak sense, a communication-avoiding technique primarily eliminates its
  negative impact---it might not reduce the actual communication.
  Such techniques are {\bf communication-flaw-avoiding}.
\end{enumerate}

\section{Communication-avoiding techniques in a strong sense}
\label{section:strong}

\begin{table}[htb]
 \caption{
   Overview of discussed communication-avoiding techniques. Some 
   labels are subjective and depend, in practice, very much on the particular
   application. However, it is fair to assume that all techniques are strong in
   a sense that they actually eliminate communication.
 }
 \setlength{\tabcolsep}{0pt}
 \begin{center}
%
  \begin{tabular}{p{4.8cm}|p{1.8cm}lllll}
  & \rotatebox{90}{Stable} 
  & \rotatebox{90}{Horizontal/Vertical} 
  & \rotatebox{90}{Generic} 
  & \rotatebox{90}{Preserve} 
  & \rotatebox{90}{Monotonous} 
  & \rotatebox{90}{Strong} \\
  \hline
  Temporary variable/storage elimination 
   & yes & H+V & no  & no  & yes & yes \\
  Recomputation  
   & yes & H   & no  & yes & yes & yes \\
  Reduced precision 
   & no  & H+V & yes & yes & yes & yes \\
  Compression
   & no(lossy)/yes &   
           H+V & yes & yes & yes & yes \\
  RLE
   & yes & H+V & yes & yes & no & yes \\
  Padding
   & yes & H+V & yes & yes & yes & yes \\
  \hline
  Overlapping (loop unrolling)
   & yes & H   & yes & yes & yes & yes \\
  Anarchic 
   & yes/no  & H+V & no  & no  & no & yes  \\
  \hline
  Tiling/serialisation 
   & yes & V   & yes & yes & yes & yes \\
  Task fusion
   & yes & V   & no  & yes & yes & yes \\
  \hline
  Sparsify collective graph (dynamically)
   & yes & H   & yes(no) & yes & yes & yes \\ 
 \end{tabular}
 \end{center}
\end{table}

\subsection{Volume reduction}
Volume reduction is the most obvious communication-avoiding technique.
A classic example is Gaussian Elimination which reorders entries such that
the number of fill-ins is reduced \cite{Demmel:2013:CA}.
Today, such algorithms are integral part of all major direct solver packages.
Algorithmic steps are permuted such that
zero entries in a data structure are not overwritten with non-zero data.
Consequently, follow-up later steps which would otherwise have to
eliminate non-zero entries can be skipped.
Yet, this is not the prime goal from a data point of view.
The prime target besides a sparsification of the execution/task graph is to
avoid a temporary blow-up of the used data structures.
The observation behind this technique is that an intelligent permutation of
computations early in the algorithm's run implies that we don't have to
store large temporary data.

If this fails, a second technique is the
replacement of temporary data/variables with recomputations.
Results are not stored persistently and thus don't have to be held or
communicated. 
Instead, we compute results on-the-fly again whenever we need them.
This technique trades memory for computations.
It is thus particularly attractive if we suffer from horizontal communication
flaws.

If it is not possible to avoid temporary data storage or data usage, precision
reduction becomes popular
\cite{Dickov:14:InfinibandCompression,Kuhn:2016:DCC,Laguna:2019:GPUMixer,Langou:2006:IterativeRefinement,Lindstrom:06:FastFloatingPointCompression}.
Machine learning pushes the introduction of precision reduction
\cite{Higham:2019:SqueezingMatrix}, but it is natural to exploit new native
hardware formats with reduced memory footprint in scientific computations, too.
Such a switch yields massive memory footprint reductions though might proof to
be problematic for the stability of the computations. 
Therefore, we currently witness massive investments into research around
algorithms which combine different precisions without compromising on the output
quality \cite{Higham:2017:News}, or tools which guide the developers through the
zoo of novel precisions.

Our own work sticks to standard IEEE precision for all computations, but
translates all data into hierarchical representations prior to their storage
\cite{Bungartz:10:Precompiler,Eckhardt:15:SPHCompression,Weinzierl:17:BoxMG}.
An on-the-fly analysis of the data yields how many significant bits (bits that
hold information) are held per variable.
We then cut down the data after a suitable
number of bytes without loosing information before we stream them out of the
core.
After the compressed, low footprint data is loaded again, we convert it back
into a native precision.
Other papers propose similar storage schemes which store only few bits per datum
relative to a (predicted) reference value
\cite{Ratanaworabhan:06:LosslessCompression}.

For integer data or sparse records holding lots of zeroes, 
run-length encoding has proven to be particularly useful.
In its simplest case, it eliminates long sequences of zeroes for the price
of additional header flags in a data stream \cite{Schreiber:13:Sfc-based}.
In particular for modern high-level programming languages, it is a CA technique
to eliminate the omnipresent padding and thus to squeeze out bytes without
semantics from the data
\cite{Bungartz:10:Precompiler}.

\subsection{Synchronisation elimination}
Volume reduction tackles bandwidth penalties. 
Modern architectures however suffer from limited bandwidth and
high latency.
Latency notably causes problems whenever an algorithm synchronises various
compute units too often.
It is non-trivial to avoid synchronisation and typically impossible to
eliminate it completely, as algorithms synchronise to keep their distributed
states consistent.

Overlapping data spaces and running calculations redundantly tackles the
synchronisation challenge.
It is a tool classic in stencil-based linear algebra.
The idea is to revisit the data decomposition.
Let the data be decomposed into chunks such that each core or node can run its
evaluations independently of the other compute units on its chunk.
We achieve this by augmenting the data chunk with halos, i.e.~with copies from
the other chunks such that the calculation itself does not need any data
transfer.
After the computation, these halos are updated, as other compute units likely
have changed their value.
We can now analyse our data access needs and make the halos bigger such that we
can do two calculations in a row. 
This typically requires us to run some calculations within the (larger) halos,
too:
we ``mirror'' some calculations done on other chunks. 
The approach can be recursively extended.
Within the overlaps, we run redundant calculations, but this 
redundancy is designed such that the redundant computations allow the
algorithm to immediately run a subsequent calculation without the need to
synchronise.
We reduce the synchronisation frequency; though at the price of an increased
data volume.
Increased data overlaps are a valid label for this technique.
The CA flavour however is emphasised once we highlight that the technique
involves loop unrolling of the outer algorithm loop and synchronises only once
per unrolled loop body.
$s$-step  Krylov methods are close to this concept
though they give up on some data consistency \cite{Ghysels:13:HideLatency}.

A radical alternative to enlarged halos which also reduces the synchronisation
is anarchic programming \cite{Wolfson:2019:AsynchronousJacobi}.
We drop the consistency considerations that led to enlarged data spaces and
ignore whether incoming information (data updates) from other entities have
already arrived.
Loops on data chunks continue to iterate, knowing that eventually another code
part will feed them with the required input data.
Such an anarchic paradigm requires the underlying algorithm to be robust
w.r.t.~inconsistent input data.
Elliptic problems seem to suit this requirement.

\subsection{Single-touch and single-load semantics}
Whenever data volume or latency pose problems, these
problems are amplified if information has to be communicated multiple times.
We rarely find distributed memory codes that exchange messages multiple times.
It happens however frequently that data is moved through the caches multiple
times (capacity misses) or that codes swap data to and from the
GPGPU due to their limited memory capacity.

The optimisation of algorithms towards particular cache architectures is
well-under\-stood.
Key ingredient here is typically proper tiling (chunkification) such that
tiles/chunks fit into the respective caches, and the attempt to increase the
spatial and temporal locality of data accesses
\cite{Kowarschik:03:CacheTechniquesOverview}.
Algorithms or loops, respectively, that 
read (array) data multiple times are reordered such that data that are required
briefly after each other are close in the main memory, and such that data that
are required at one point are re-used again briefly thereafter.
We adapt the data access pattern of an algorithm and the data storage to each
other.
Sophisticated cache blocking techniques
\cite{Kowarschik:03:CacheTechniquesOverview,Saxena:2017:CacheAwareDD} exploit that most memory is
organised in pages/blocks/lines, i.e.~architectures do not exchange single entities of data but always transfer whole segments of memory---a
built-in message agglomeration.
They make the chunk sizes match particular cache sizes.

Clever data access and data storage reordering reduces multiple loads into
caches.
An extreme case of such clever reordering is the total serialisation of an
algorithm \cite{Weinzierl:11:Peano}. 
Data are reordered such that the algorithm does not require any
indirect memory accesses anymore, but reads all data in as stream and holds
data temporarily only in stacks, streams or small arrays.
As a consequence, an algorithm becomes inherently cache-optimal (cache
oblivious)---the probability that the head of a stack remains in a cache is
always very high as long as the number of stacks remains small---and avoids
communication over the memory interconnects.

Data ordering and operation orchestration go hand in hand. 
On the orchestration side, task fusion is the most important single-touch
optimisation \cite{Charrier:2019:StopTalkingToMe}.
Task fusion reorders operations (tasks) such that two tasks that access the same
piece of data are evaluated directly after another. 
One might thus (logically) fuse the two pieces of codes into one big task.
If tasks operate on a reasonably small amount of data, we may assume that all
data are held in fast caches after the termination of the first task.  
Task fusion is the task language's counterpart to loop fusion on arrays from a
CA point of view.

\subsection{Localisation}
Bandwidth and latency problems require different actions.
Quantitatively, these problems are affected by the number of participating
entities:
the more partners involved in communication, the harder the challenge.
In many cases, a reasonably small number of communication partners makes
communication flaws hide behind runtime noise.
If problems however do arise, it is sometimes only few communication partners
that cause delays on all other entities.
In such cases, it is reasonable to break up many-to-many data exchange
(collectives).
Rather than global data exchange, we localise data flow.
Such a sparsification of the data exchange graph can be done a priori by
analysing data dependencies.
In multiscale algorithms for example, it might be sufficient to let a
computation depend only on a certain subregion of one scale rather than the
whole level of one scale \cite{Mitchell:2019:AMG-DD}.
Sparse collectives have been proposed for MPI
\cite{Hoefler:09:SparseCollectives}, though one can always implement such sparse
graphs via tailored MPI (sub-)communicators.

Instead of a static localisation, we have studied algorithms which localise the
data exchange graph dynamically \cite{Weinzierl:15:PIC}.
The concept here is that we query the algorithm for predictions where
collectives actually transfer information.
If collectives exchange data along a sparse communication graph (many entries
are zero, e.g.), it is again reasonable to replace the collective with a set of
point-to-point operations.

\section{Communication-avoiding techniques in a weak sense}
\label{section:weak}

\begin{table}[htb]
 \caption{
   Further communication-avoiding techniques in a weak sense, i.e.~they do not
   actually eliminate communication but they tackle communication's negative
   impact.
 }
 \setlength{\tabcolsep}{0pt}
 \begin{center}
%
  \begin{tabular}{p{4.8cm}|p{1.8cm}lllll}
  & \rotatebox{90}{Stable} 
  & \rotatebox{90}{Horizontal/Vertical} 
  & \rotatebox{90}{Generic} 
  & \rotatebox{90}{Preserve} 
  & \rotatebox{90}{Monotonous} 
  & \rotatebox{90}{Strong} \\
  \hline
  Non-blocking exchange 
   & yes & H & yes & yes & yes & no \\
  Prefetching
   & yes & V & yes & yes & yes & no \\
  \hline
  Optimistic 
   & yes & H   & no  & yes & no & no  \\
  Pipelining
   & no  & H   & no  & yes & yes & no \\
  \hline
  Homogenisation
   & yes & H+V & yes & yes & no  & yes
 \end{tabular}
 \end{center}
\end{table}

\subsection{Data movement hiding}
The hardware's organisation into levels (vertical) and nodes
(horizontal) introduces hardware asynchronicity since the entities can move around
data independently and, in particular, while core components compute.
This allows us to hide data movement cost.
Hiding is an example of a weaker interpretation of CA, as the
communication is not literally avoided yet negative impact is.

On the distributed memory side, non-blocking data exchange means that we trigger
data transfer for a particular memory region, continue to compute, and later
check whether the transfer has completed.
This is a rather old feature in MPI, though still sometimes challenging to
use/realise
\cite{Hoefler:08:SacrificeThread,Sergent:2018:OverlapCommunicationComputation,Wittmann:13:AsynchMPI}.
Within a node, all caches today have built-in prediction features to
transfer data from and to caches.
This also holds for new deep memory
\cite{Boyandin:18:OptaneInMemoryDB,Charrier:19:EnergyAndDeepMemory,Kudryavtsev:17:NUMAOptane}.
Agnostic of the exact caching strategies, least recently used
(LRU) seems to describe most cache behaviour reasonably.
LRU describes reactive cache behaviour: 
the ``oldest'' data is removed from the cache if space is required.  
Explicit prefetching can guide the caching as it triggers
data movements ahead of time.
It hands a pro-active technique over to developers.

\subsection{Synchronisation hiding}

Optimism is a powerful communication-avoiding technique to hide
synchronisation.
Optimistic algorithms synchronise program parts, but the receiver side does
not wait for the synchronisation input.
Instead, it optimistically assumes what information will drop in.
If this assumption turns out to be wrong later during the computation, the
receiver side either rolls back and redoes its computations with the actual
values \cite{Charrier:2019:StopTalkingToMe} or takes the actual data into account at the earliest
next opportunity.
If rollbacks are required, optimistic codes' runtime might deteriorate unless
we can immediately stop all calculations once invalidating data drops in.

Pipelining \cite{Ghysels:13:HideLatency,Ghysels:14:HideSynchronisation}
applies the data overlap idea to the control flow.
Its idea is that we bring all the computations forward that feed
into computations on another compute entity.
In turn, we receive data early that we need locally to determine results
required in later stages.
Data required for the follow-up step is already ``in the pipeline'' while we
continue to compute.
To construct the required level of concurrency,
pipelining is often combined with redundant calculations.
Entities exchange data feeding into critical calculations early, but 
each rank then computes the results independently from the others. 
Both pipelining and increased overlaps from Section
\ref{section:strong} contradict CA's objective to keep the memory footprint
small.
Many approaches relying on redundant computations furthermore become sensitive
to stability issues.

\subsection{Homogenisation}
Computers tend to be underequipped with memory access speed and with network
speed.
Yet, memory controllers and network cards  usually are sufficiently fast to
handle a few cores or nodes, respectively.
Service requests from everybody at once however make them struggle.
A weak communication-avoiding technique to address this flaw is
to homogenise bandwidth and network requests over time:
we sprinkle communication steps over the actual execution.
In particular, we break up codes written in in terms of compute
phases vs.~communication phases which take turns and make many (smaller)
communication steps interrupt the computation.
On the node, the same pattern implies that we carefully orchestrate all tasks
such that compute-intense tasks and memory-intense tasks take turns
\cite{Charrier:2019:Enclave,Charrier:2019:StopTalkingToMe}.
The goal here is to ensure that some compute facilities (cores) exploit all
available memory bandwidth, while the others spend the majority of their time in
actual computations and do not interfere.

\section{Conclusion}
\label{section:conclusion}

Communication-avoiding techniques have to become omnipresent if the current
hardware evolution trend endures.
The present collection of CA techniques is a rather arbitrary set inspired by
the author's own work and selected examples known to the author.
It is certainly not comprehensive.
The present classification expresses the author's personal opinion.
It is likely neither waterproof nor comprehensive either.
Yet, the present collection plus its classification can be a starting point to
think about CA techniques in a systematic way covering whole application
landscapes.
Eventually, a concerted push by the whole community is required to establish a more
complete catalogue of these techniques similar to design patterns
\cite{Gamma:94:DesignPatterns}.
This way, future generations of scientists will be able to rely on a
well-documented and formalised knowledge base.
Without such an effort, we run risk to be left behind with hardware that we cannot
(efficiently) use.

An inherent obstacle towards the adoption of communication-avoiding techniques
is the fact that they often require large code refactorings or even rewrites
(cmp.~preserve characterisation in Section \ref{section:terminology}); at a time
where researchers advocate to switch to black-box software building blocks to make software development more mature, efficient and sustainable.
Machine learning serves as prime example for the latter.
It champions the usage of established software frameworks
within multiple research projects.
Our discussion highlights however that a na\"ive assembly of software blocks
representing an algorithm's steps (data-flow paradigm) might be an ill-fit to
future machines:
If we complete one computing step, dump or stream the outcome into the next
algorithmic phase, and then continue, we will struggle to use
communication-avoiding techniques.

It is almost sarcastic that the discipline that pushed us into petascale, linear
algebra, and the discipline which drove the hardware development
towards the pre-exascale era, i.e.~machine learning, are kind of agnostic or
even ``contraproductive'' towards this particular challenge flavour.
One reason for the success of large-scale linear algebra packages is
without a doubt their clear encapsulation.
Data transfer is a challenge picked up within the software, but the API is not
designed towards CA\footnote{The author is well-aware that PETSc, e.g., offers a
matrix-free usage mode. Yet, it seems not to be the most popular way to use
the software within complex PDE discretisations; at least the author is not
aware of extended documentation and examples how to integrate it into an
application landscape outside of the PETSc sanctum.}.
One reason for the success of machine learning frameworks is the fact that we
can use them almost as black-box. 
This implies that users tend to realise a data pipeline architecture where data
transfer is not a high priority design quality criterion.
The visualisation community in contrast is well-aware of this rising
challenge---mainly due to the large IO penalties computers already face---and
pushes in-situ postprocessing and software that is made for in-situ processing.
The computing community either has to follow or establish usable
refactoring techniques \cite{Fowler:99:Refactoring} such that developers can
start from black-box units and then refactor towards CA.

Finally, it is important for the scientific computing community to continue to
push vendors towards real co-design. 
At the moment, engineering constraints and economic interest (machine learning)
seem to steer the direction into which our hardware evolves.
While this evolution stimulates the development of communication-avoiding
techniques and thus stimulates an area of research, it is important to challenge
vendors over and over again.
Would it not be possible to build better-balanced systems?

\acknowledgments
The author expresses his thanks to all collaborators and students
that helped him to study various flavours of communication-avoiding algorithms.
They are enlisted as co-authors on the cited papers.
A lot of the author's recent communication-avoiding research has been conducted
under the umbrella of the ExaHyPE project supported by the European Union’s Horizon 2020
research and innovation programme under grant agreement No 671698 (ExaHyPE).
The ideas around the particle handling, i.e.~the break up of collectives,
evolved from a research visit of and collaboration with Bart Verleye who had
been funded by Intel and by the Institute for the Promotion of Innovation through Science and Technology in Flanders (IWT).
Single-touch algorithm rewrites in the context of additive multigrid has been
kicked off by a collaboration with Bram Reps who had been supported by the
FP7/2007-2013 programme under grant agreement No 610741 (EXA2CT). 
Volume compression concepts for stencil-based multigrid solvers arose from a
collaboration with Marion Weinzierl who received support from the
International Graduate School of Science and Engineering (IGSSE), Technische
Universit\"at M\"unchen.

Special thanks are due to Marco Paganoni and Michele Fumagalli for establishing
the contact and eventually inviting me to the Big Data Analytics workshop in
Varenna in June 2019. This manuscript picks up some aspects of my talk there.
Special thanks are also due to Marion Weinzierl who carefully proofread and
commented on the present manuscript.

\bibliographystyle{plain}
\bibliography{paper}  

\end{document}